\documentstyle[12pt]{article}
\newcommand{\cl}{\centerline}
\renewcommand{\theequation}{\arabic{equation}}
\newcommand\beq{\begin{equation}}
\newcommand\eeq{\end{equation}}
\newcommand\bea{\begin{eqnarray}}
\newcommand\eea{\end{eqnarray}}

\begin{document}

\begin{titlepage}
\setlength{\textwidth}{5.0in}
\setlength{\textheight}{7.5in}
\setlength{\parskip}{0.0in}
\setlength{\baselineskip}{18.2pt}
\begin{center}
{\large{\bf BRST symmetries in SU(3) linear sigma model}}\par
\vskip 0.5cm
\begin{center}
{Soon-Tae Hong$^{1,2}$ and Su Houng Lee$^{2}$}\par
\end{center}
\vskip 0.4cm
\begin{center}
{$^{1}$Department of Physics and Basic Science Research Institute,}\par
{Sogang University, C.P.O. Box 1142, Seoul 100-611, Korea}\par
\vskip 0.5cm
{$^{2}$Department of Physics and Institute of Physics and Applied Physics,}\par
{Yonsei University, Seoul 120-749, Korea}\par
\end{center}
\vskip 0.3cm
\cl{November 16, 2001}
%\cl{\today}
\vskip 0.5cm
\vfill
\begin{center}
{\bf ABSTRACT}
\end{center}
\begin{quotation}

We study the BRST symmetries in the SU(3) linear sigma model which is constructed 
through introduction of a novel matrix for the Goldstone boson fields satisfying 
geometrical constraints embedded in SU(2) subgroup.  To treat 
these constraints we exploit the improved Dirac quantization scheme.  We also 
discuss phenomenological aspacts in the mean field approach to this model.      

\vskip 0.5cm
\noindent
%---------------------------------------------------------------------\\
\vskip 0.5cm
\noindent
\end{quotation}
\end{center}
\end{titlepage}

\newpage

\section{Introduction}

Nowadays there have been considerable discussions concerning the strangeness 
in hadron physics.  Recently, the SAMPLE experiment~\cite{sample01} 
reported the proton's neutral weak magnetic form factor, which has been suggested 
by the neutral weak magnetic moment measurement through parity violating electron
scattering\cite{mck89}.  In fact, the SAMPLE Collaboration obtained the positive 
experimental data for the proton strange magnetic form factor~\cite{sample01} 
$G_M^s (Q^2 = 0.1 {\rm (GeV/c)}^2) = +0.14 \pm 0.29~{\rm (stat)} 
\pm 0.31~{\rm (sys)}$.  This positive experimental value is contrary to the 
negative values of the proton strange form factor which result from most of the model 
calculations except the predictions~\cite{hong97} based on the SU(3) chiral bag model~\cite{gerry791} 
and the recent predictions of the chiral quark 
soliton model~\cite{kim298} and the heavy baryon chiral perturbation theory~\cite{meissner00}.  
(See Ref.~\cite{hongpr01} for more details.)  

It is also well known in the strangeness hadron physics that kaon 
condensation~\cite{kaplan86,gerry87,kaoncon,schafer00prd}in nuclear matter of densities may have an impact on the formation of low mass black holes 
instead of neutron stars for masses on the order of 1.5 solar masses~\cite{gerry94}.  
Beginning with the proposal of 
kaon condensation in 1986~\cite{kaplan86}, the theory of kaon condensation in dense matter has become one of 
the central issues in nuclear physics and astrophysics together with the supernova collapse.  The $K^{-}$ condensation at a few times nuclear matter density was later interpreted~\cite{gerry87} in terms of cleaning of $\bar{\rm q}$q condensates from 
the quantum chromodynamics vacuum by a dense nuclear matter and also in terms 
of phenological off-shell meson-nucleon interactions~\cite{yabu93}.  Recently, the kaon condensation was revisited in the context of the color superconductivity in color-flavor phase~\cite{schafer00prd}

On the other hand, the Dirac method \cite{di} is a well known formalism to 
quantize physical systems with constraints.  In this method, the Poisson brackets 
in a second-class constraint system are converted into Dirac brackets to attain 
self-consistency.  The Dirac quantization scheme has been also applied to the nuclear phenomenology~\cite{sulee88prd,sulee88plb}.  The Dirac brackets, however, are generically 
field-dependent, nonlocal and contain problems related to ordering of field 
operators.  These features are unfavorable for finding canonically conjugate 
pairs. However, if a first-class constraint system can be constructed, one can 
avoid introducing the Dirac brackets and can instead use Poisson brackets to 
arrive at the corresponding quantum commutators.  To overcome the above problems, 
Batalin, Fradkin, and Tyutin~\cite{BFT} developed a method which converts the 
second-class constraints into first-class ones by introducing auxiliary fields.  
Recently, this improved Dirac formalism has been applied to several models of current 
interest. (See Ref.~\cite{hongpr01} for more details.)  In particular, the relation 
between the standard Dirac and improved Dirac schemes was clarified in the SU(2) 
Skyrmion model by introducing additional correction terms~\cite{sk2}.  Recently, the 
improved Dirac Hamiltonian method was also applied to the SU(2) and SU(3) Skyrmion 
models~\cite{su2bft,su3bft} to construct a first-class Hamiltonian and the BRST symmetries, and 
to the superqualiton model~\cite{hong01cfl} to investigate the color superconductivity 
in color-flavor phase.  Moreover, the BRST symmetry~\cite{brst} has been 
studied~\cite{su2bft} in the SU(2) Skyrmion in the framework of the BFV 
formalism~\cite{fradkin75} which is applicable to theories 
with the first class constraints by introducing canonical sets of ghosts and 
anti-ghosts together with auxiliary fields.  The BRST symmetry can be also 
constructed by using the residual gauge symmetry interpretation of the BRST 
invariance~\cite{yee93}.

The motivation of this paper is to systematically apply the improved Dirac 
scheme~\cite{BFT,hongpr01} to the SU(3) linear sigma model to construct the 
BRST symmetries in this phenomenological model.  In section 2 we construct 
the SU(3) linear sigma model by introducing a novel matrix for the Goldstone bosons 
which satisfy geometrical second-class constraints.  To treat these 
constraints we exploit the improved Dirac scheme to convert the second-class 
constraints into first-class ones. In section 3 we construct first-class 
physical fields and directly derive the compact expression of a first-class 
Hamiltonian in terms of these fields.  We construct in section 4 the BRST 
invariant effective Lagrangian and its corresponding BRST transformation rules in 
the SU(3) linear sigma model.  In section 5 we discuss the phenomenology 
of the pion and kaon condensates. 

%%%%%%%%%%%%%%%%%%%%%%%%%%%%%%%%%%%%%%%%%%%%%%%%%%%%%%%%%%%%%%%%%%%%%%%%%%

\section{Model and constraints}
\setcounter{equation}{0}
\renewcommand{\theequation}{\arabic{section}.\arabic{equation}}

%%%%%%%%%%%%%%%%%%%%%%%%%%%%%%%%%%%%%%%%%%%%%%%%%%%%%%%%%%%%%%%%%%%%%%%%%%

In this section we apply the improved Dirac scheme~\cite{BFT,hongpr01} to the 
SU(3) linear sigma model, which is a second-class constraint system.  
We start with the SU(3) linear sigma model Lagrangian of the form 
\bea
L&=&\int {\rm d}^{3}x\left[\frac{1}{2}{\rm tr}(\partial_{\mu}M\partial^{\mu}M^{\dagger})
-\frac{1}{2}\mu_{0}^{2}{\rm tr}(MM^{\dagger})-\frac{1}{4}\lambda_{0}
[{\rm tr}(MM^{\dagger})]^{2}\right.\nonumber\\
& &\left.+\bar{\psi}i\gamma^{\mu}\partial_{\mu}\psi
-g_{0}(\bar{\psi}_{L}M\psi_{R}+\bar{\psi}_{R}M^{\dagger}\psi_{L})
\right]
\label{lagsu3}
\eea
where we have introduced a novel matrix for the Goldstone bosons satisfying geometrical second-class constraints~\footnote{In the previous 
works~\cite{pagels75}, they have used a different ansatz for $M$ such as 
$M=\sum_{i=0}^{8}(\sigma_{i}^{\prime}+i\pi_{i})\lambda_{i}/\sqrt{2}$ with 
nonets of scalar $\sigma_{i}^{\prime}$ and pseudoscalar fields $\pi_{i}$ which
transform according to the $3\otimes\bar{3}+\bar{3}\otimes 3$ representation of SU(3)$\times$SU(3).}
\beq
M=\frac{1}{\sqrt{2}}(\sigma\lambda_{0}+i\pi_{a}\lambda_{a}),~~~a=1,\cdots,8,
\eeq
with $\lambda_{0}=\sqrt{\frac{2}{3}}I$ ($I$: identity) and the Gell-Mann matrices 
normalized to satisfy $\lambda_{a}\lambda_{b}=\frac{2}{3}\delta_{ab}
+(if_{abc}+d_{abc})\lambda_{c}$.  Here we have meson fields $\pi_{a}=
(\pi_{i},\pi_{M},\pi_{8})$ with $\pi_{i}$, $\pi_{M}$ and $\pi_{8}$ being the pion, kaon and eta fields, respectively, and the chiral fields $\psi_{L}$ and $\psi_{R}$ defined as
\beq
\psi_{R,L}=\frac{1\pm\gamma_{5}}{2}\psi.
\eeq
Note that we have used the SU(3) linear sigma model with U(3)$\times$U(3) group structure so that we could incorporate the $\sigma$ field consistently, as in the chiral bag model~\cite{eta97}. 

The Lagrangian (\ref{lagsu3}) can then 
be rewritten in terms of the meson fields $\pi_{a}$ as follows
\bea
L&=&\int {\rm d}^{3}x\left[\frac{1}{2}\left(\partial_{\mu}\sigma\partial^{\mu}\sigma
+\partial_{\mu}\pi_{a}\partial^{\mu}\pi_{a}\right)
-\frac12\mu_{0}^{2}(\sigma^{2}+\pi_{a}\pi_{a})-\frac{1}{4}\lambda_{0}
(\sigma^{2}+\pi_{a}\pi_{a})^{2}\right.\nonumber\\
& &\left.+\bar{\psi}i\gamma^{\mu}\partial_{\mu}\psi
-g_{0}\bar{\psi}\frac{1}{\sqrt{2}}(\sigma+i\gamma_{5}\pi_{a}\lambda_{a})\psi
\right]
\label{lagsu32}
\eea
where we have assumed the SU(3) flavor symmetry for simplicity.  Here the sigma 
and pion fields $(\sigma$, $\pi_{i})$ are constrained to satisfy 
the geometric constraints on the SU(2) subgroup manifold
\begin{equation}
\Omega_{1}=\sigma^{2}+\pi_{i}\pi_{i}-f_{\pi}^{2}\approx 0.  \label{c1}
\end{equation}
By performing the Legendre transformation, one can obtain the canonical 
Hamiltonian, 
\bea
H_c&=&\int{\rm d}^{3}x\left[\frac12\left(\pi_{\sigma}^{2}+\pi_{\pi}^{a}\pi_{\pi}^{a}
\right)
+\frac12((\partial_{i}\sigma)^{2}+(\partial_{i}\pi_{a})^{2})
+\frac12\mu_{0}^{2}(\sigma^{2}+\pi_{a}\pi_{a})\right.\nonumber\\
& &\left.
+\frac{1}{4}\lambda_{0}(\sigma^{2}+\pi_{a}\pi_{a})^{2}
+\bar{\psi}i\gamma^{i}\partial_{i}\psi
+g_{0}\bar{\psi}\frac{1}{\sqrt{2}}(\sigma+i\gamma_{5}\pi_{a}\lambda_{a})\psi\right]
\label{hc}
\eea
where $\pi_{\sigma}$ and $\pi_{\pi}^{a}$ are the canonical momenta 
conjugate to the fields $\sigma$ and $\pi_{a}$, respectively, 
given by 
\beq
\pi_{\sigma}=\dot{\sigma},~~~~~\pi_{\pi}^{a}=\dot{\pi}_{a}
\label{cojm}
\eeq
and we have used $\bar{\psi}$ for the canonical momenta conjugate to the 
fields $\psi$ instead of $\pi_{\psi}^{\dagger}=i\psi^{\dagger}$ for simplicity. 

Now we want to construct Noether currents under the SU(3)$_{L}\times$SU(3)$_{R}$ local group transformation.  Under infinitesimal isospin transformation in the 
SU(3) flavor channel~\cite{hongpr01}
\bea
\psi\rightarrow \psi^{\prime}&=&(1-i\epsilon_{a}\hat{Q}_{a})\psi,\nonumber\\
M\rightarrow M^{\prime}&=&(1-i\epsilon_{a}\hat{Q}_{a})M
(1+i\epsilon_{a}\hat{Q}_{a}),
\eea
where $\epsilon_{a}(x)$ the local angle parameters of the group transformation
and $\hat{Q}_{a}=\lambda_{a}/2$ are the SU(3) flavor charge operators given by 
the generators of the symmetry, the Noether theorem yields the {\it conserved} 
flavor octet vector currents (FOVC) for the Lagrangian (\ref{lagsu32})
\beq
J_{V}^{\mu a}=\bar{\psi}\gamma^{\mu}\hat{Q}_{a}\psi
+\frac{i}{2}{\rm tr}\left([M,\hat{Q}_{a}]\partial^{\mu}M^{\dagger}
+\partial^{\mu}M [M^{\dagger},\hat{Q}_{a}]\right).
\label{jvmua}
\eeq
In addition one can see that the electromagnetic (EM) currents $J_{EM}^{\mu}$ 
can be easily constructed by replacing the SU(3) flavor charge operators 
$\hat{Q}_{a}$ with the EM charge operator $\hat{Q}_{EM}=\hat{Q}_{3}+\frac{1}
{\sqrt{3}}\hat{Q}_{8}$ in the FOVC (\ref{jvmua}).  Moreover one can 
obtain the charge density $\rho$ as follows
\beq
\rho=\psi^{\dagger}\hat{Q}_{EM}\psi
+(f_{3ab}+\frac{1}{\sqrt{3}}f_{8ab})\pi_{a}\pi_{\pi}^{b}.
\label{jvmuem}
\eeq

Now we introduce the chemical potentials $\mu=(\mu_{\pi},\mu_{K})$ corresponding to the charge densities $\rho=(\rho_{\pi},\rho_{K})$ in the pion and kaon flavor 
channels to yield the Hamiltonian in the kaon condensed matter
\beq
H=H_{c}+\mu_{\pi}\rho_{\pi}+\mu_{K}\rho_{K}
\label{hamdense}
\eeq
where the charge densities are now explicitly given as 
\bea
\rho_{\pi}&=&\psi^{\dagger}(\hat{Q}_{u}+\hat{Q}_{d})\psi
+\pi_{1}\pi_{\pi}^{2}-\pi_{2}\pi_{\pi}^{1},\nonumber\\
\rho_{K}&=&\psi^{\dagger}\hat{Q}_{s}\psi
+\pi_{4}\pi_{\pi}^{5}-\pi_{5}\pi_{\pi}^{4},
\label{rhoq}
\eea
with $\hat{Q}_{q}$ being the q-flavor EM charge operators.  Here we have ignored the 
beta equilibrium for simplicity.  Note that by defining the flavor projection operators
\beq
\hat{P}_{u,d}=\frac{1}{3}\pm\frac12 \lambda_{3}+\frac{1}{2\sqrt{3}}\lambda_{8},~~~
\hat{P}_{s}=\frac{1}{3}-\frac{1}{2\sqrt{3}}\lambda_{8},
\label{projectionop}
\eeq
satisfying $\hat{P}_{q}^{2}=\hat{P}_{q}$ and $\sum_{q}\hat{P}_{q}=1$, one can 
easily construct the q-flavor EM charge operators $\hat{Q}_{q}=\hat{Q}_{EM}
\hat{P}_{q}=Q_{q}\hat{P}_{q}$.  

On the other hand, the time evolution of the constraint $\Omega_1$ yields 
an additional secondary constraint 
\begin{equation}
\Omega_{2}=\sigma\pi_{\sigma}+\pi_{i}\pi_{\pi}^{i}\approx 0,  
\label{const22}
\end{equation}
and $\Omega_{1}$ and $\Omega_{2}$ form a second-class constraint algebra 
\begin{equation}
\Delta_{kk^{\prime}}(x,y)=\{\Omega_{k}(x),\Omega_{k^{\prime}}(y)\}
=2\epsilon^{kk^{\prime}}(\sigma^{2}+\pi_{i}\pi_{i})\delta(x-y)  \label{delta}
\end{equation}
with $\epsilon^{12}=-\epsilon^{21}=1$.

Using the Dirac brackets \cite{di} defined as
\beq
\{A(x),B(y)\}_{D}=\{A(x),B(y)\}-\int d^3z d^3 z^{\prime}
\{A(x),\Omega_{k}(z)\}\Delta^{k k^{\prime}}
\{\Omega_{k^{\prime}}(z^{\prime}),B(y)\}
\end{equation}
with $\Delta^{k k^{\prime}}$ being the inverse of $\Delta_{k k^{\prime}}$ in
Eq. (\ref{delta}), we obtain the following commutators   
\begin{eqnarray}
\{\sigma(x),\sigma(y)\}_{D}&=&\{\pi_{\sigma}(x),\pi_{\sigma}(y)\}_{D}=0,
\nonumber\\
\{\sigma(x),\pi_{\sigma}(y)\}_{D}&=&\left(1-\frac{\sigma^{2}}{\sigma^{2}
+\pi_{k}\pi_{k}}\right)\delta(x-y),  
\nonumber\\
\{\pi_{a}(x),\pi_{b}(y)\}_{D}&=&0,  \nonumber\\
\{\pi_{a}(x),\pi_{\pi}^{b}(y)\}_{D}&=&
\left(\delta_{ab}-\frac{\pi_{i}\pi_{j}}{\sigma^{2}
+\pi_{k}\pi_{k}}\delta_{ai}\delta_{bj}\right)\delta(x-y),  
\nonumber\\
\{\pi_{\pi}^{a}(x),\pi_{\pi}^{b}(y)\}_{D}&=&
\frac{1}{\sigma^{2}+\pi_{k}\pi_{k}}\left(\pi_{j}\pi_{\pi}^{i}
-\pi_{i}\pi_{\pi}^{j}\right)\delta_{ai}\delta_{bj}\delta(x-y),  
\nonumber\\
\{\psi(x),\psi(y)\}_{D}&=&
\{\pi_{\psi}^{\dagger}(x),\pi_{\psi}^{\dagger}(y)\}_{D}=0,
\nonumber\\
\{\psi(x),\pi_{\psi}^{\dagger}(y)\}_{D}&=&\delta(x-y).
\label{commdirac}
\end{eqnarray}

Following the improved Dirac formalism~\cite{BFT,hongpr01} which systematically
converts the second-class constraints into first-class ones, we introduce
two auxiliary fields $(\theta,\pi_{\theta})$ with the Poisson brackets 
\begin{equation}
\{\theta (x), \pi_{\theta}(y)\}=\epsilon^{ij}\delta (x-y).  \label{phii}
\end{equation}
The first class constraints $\tilde{\Omega}_{i}$ are then constructed as 
\beq
\tilde{\Omega}_{1}=\Omega_{1}+2\theta,~~~
\tilde{\Omega}_{2}=\Omega_{2}-(\sigma^{2}+\pi_{k}\pi_{k})\pi_{\theta},  
\label{1stconst}
\eeq
which satisfy the closed algebra 
$\{\tilde{\Omega}_{i}(x),\tilde{\Omega}_{j}(y)\}=0$.

%%%%%%%%%%%%%%%%%%%%%%%%%%%%%%%%%%%%%%%%%%%%%%%%%%%%%%%%%%%%%%%%%%%%%%%%%%

\section{First-class Hamiltonian}
\setcounter{equation}{0}
\renewcommand{\theequation}{\arabic{section}.\arabic{equation}}

%%%%%%%%%%%%%%%%%%%%%%%%%%%%%%%%%%%%%%%%%%%%%%%%%%%%%%%%%%%%%%%%%%%%%%%%%%

Now, following the improved Dirac formalism~\cite{BFT,hongpr01}, we construct the 
first-class physical fields $\tilde{{\cal F}}=(\tilde{\sigma},
\tilde{\pi_{a}},\tilde{\psi},\tilde{\pi}_{\sigma},\tilde{\pi}_{\pi}^{a},
\tilde{\pi}_{\psi})$ corresponding to the original fields
${\cal F}=(\sigma,\pi_{a},\psi,\pi_{\sigma},\pi_{\pi}^{a},\pi_{\psi})$.  
The $\tilde{{\cal F}}$'s, which reside 
in the extended phase space, are obtained as a power series in the auxiliary 
fields $(\theta,\pi_{\theta})$ by demanding that they are strongly involutive: 
$\{\tilde{\Omega}_{i}, \tilde{{\cal F}}\}=0$. 
After some lengthy algebra, we obtain the first class physical fields as 
\begin{eqnarray}
\tilde{\sigma}&=&\sigma\left(\frac{\sigma^{2}+\pi_{k}\pi_{k}+2\theta}
{\sigma^{2}+\pi_{k}\pi_{k}}\right)^{1/2},  \nonumber \\
\tilde{\pi_{i}}&=&\pi_{i}\left(\frac{\sigma^{2}+\pi_{k}\pi_{k}+2\theta}
{\sigma^{2}+\pi_{k}\pi_{k}}\right)^{1/2},  \nonumber \\
\tilde{\pi}_{\sigma}&=&(\pi_{\sigma}-\sigma\pi_{\theta})
\left(\frac{\sigma^{2}+\pi_{k}\pi_{k}}{\sigma^{2}+\pi_{k}\pi_{k}
+2\theta}\right)^{1/2},  \nonumber \\
\tilde{\pi}_{\pi}^{i}&=&(\pi_{\pi}^{i}-\pi_{i}\pi_{\theta})
\left(\frac{\sigma^{2}+\pi_{k}\pi_{k}}{\sigma^{2}+\pi_{k}\pi_{k}
+2\theta}\right)^{1/2},  \nonumber \\
\tilde{\pi}_{\bar{a}}&=&\pi_{\bar{a}},~~~
\tilde{\pi}_{\pi}^{\bar{a}}=\pi_{\pi}^{\bar{a}},~~~
\tilde{\psi}=\psi,~~~
\tilde{\pi}_{\psi}^{\dagger}=\pi_{\psi}^{\dagger},
\label{pitilde}
\end{eqnarray}
with the new notation $\bar{a}=4,5,6,7,8$.

Since any functional of the first class fields $\tilde{{\cal F}}$ is also first class, 
we can construct a first-class Hamiltonian in terms of the above first-class physical 
variables as follows  
\bea
\tilde{H}&=&\int{\rm d}^{3}x\left[\frac12\left(\tilde{\pi}_{\sigma}^{2}
+\tilde{\pi}_{\pi}^{a}\tilde{\pi}_{\pi}^{a}\right)
+\frac12((\partial_{i}\tilde{\sigma})^{2}+(\partial_{i}\tilde{\pi}_{a})^{2})
+\frac12\mu_{0}^{2}(\tilde{\sigma}^{2}+\tilde{\pi}_{a}\tilde{\pi}_{a})\right.\nonumber\\
& &\left.
+\frac{1}{4}\lambda_{0}(\tilde{\sigma}^{2}+\tilde{\pi}_{a}\tilde{\pi}_{a})^{2}
+\mu_{\pi}\tilde{\rho}_{\pi}+\mu_{K}\tilde{\rho}_{K}
+\tilde{\bar{\psi}}i\gamma^{i}\partial_{i}\tilde{\psi}
\right.\nonumber\\
& &\left.+g_{0}\tilde{\bar{\psi}}\frac{1}{\sqrt{2}}
(\tilde{\sigma}+i\gamma_{5}\tilde{\pi}_{a}\lambda_{a})\tilde{\psi}
\right].
\label{htilde}
\eea 

We then directly rewrite this Hamiltonian in terms of the original as well as 
auxiliary fields\footnote{In deriving the first class Hamiltonian $\tilde{H}$ 
of Eq. (\ref{hct}), we have used the conformal map condition, $\sigma\partial_{i}\sigma
+\pi_{k}\partial_{i}\pi_{k}=0$.} to obtain
\begin{eqnarray}
\tilde{H}&=&\int {\rm d}^{3}x~\left[
\frac12\left(\left(\pi_{\sigma}-\sigma\pi_{\theta}\right)^{2}
+\left(\pi_{\pi}^{i}-\pi_{i}\pi_{\theta}\right)^{2}\right)
\frac{\sigma^{2}+\pi_{k}\pi_{k}}
{\sigma^{2}+\pi_{k}\pi_{k}+2\theta}+\frac{1}{2}\pi_{\pi}^{\bar{a}}\pi_{\pi}^{\bar{a}}
\right.\nonumber\\
& &\left.
+\frac12((\partial_{i}\sigma)^{2}+(\partial_{i}\pi_{k})^{2}
+\mu_{0}^{2}(\sigma^{2}+\pi_{i}\pi_{i}))
\frac{\sigma^{2}+\pi_{k}\pi_{k}+2\theta}
{\sigma^{2}+\pi_{k}\pi_{k}}+\frac{1}{2}(\partial_{i}\pi_{\bar{a}})^{2}
\right.\nonumber\\
& &\left.+\frac{1}{2}\mu_{0}^{2}\pi_{\bar{a}}\pi_{\bar{a}}
+\frac{1}{4}\lambda_{0}(\sigma^{2}+\pi_{i}\pi_{i})^{2}
\left(\frac{\sigma^{2}+\pi_{k}\pi_{k}+2\theta}
{\sigma^{2}+\pi_{k}\pi_{k}}\right)^{2}+\frac{1}{4}\lambda_{0}(\pi_{\bar{a}}\pi_{\bar{a}})^{2}
\right.\nonumber\\
& &\left.+\frac{1}{2}\lambda_{0}\pi_{\bar{a}}\pi_{\bar{a}}(\sigma^{2}+\pi_{i}\pi_{i}+2\theta)
+\mu_{\pi}\rho_{\pi}+\mu_{K}\rho_{K}+\bar{\psi}i\gamma^{i}\partial_{i}\psi
\right.\nonumber\\
& &\left.+g_{0}\bar{\psi}\frac{1}{\sqrt{2}}(\sigma+i\gamma_{5}\pi_{i}\tau_{i})\psi
\left(\frac{\sigma^{2}+\pi_{k}\pi_{k}+2\theta}{\sigma^{2}+\pi_{k}\pi_{k}}\right)^{1/2}
+g_{0}\bar{\psi}\frac{1}{\sqrt{2}}i\gamma_{5}\pi_{\bar{a}}\lambda_{\bar{a}}\psi
\right],
\nonumber\\
\label{hct}
\end{eqnarray}
where we observe that the forms of the first two terms in this Hamiltonian are 
exactly the same as those of the O(3) nonlinear sigma model~\cite{o3}.  

Now we notice that, even though $\tilde{H}$ is strongly involutive with the first 
class constraints $\{\tilde{\Omega}_{i},\tilde{H}\}=0$, it does not naturally 
generate the first-class Gauss law constraint from the time evolution of the 
constraint $\tilde{\Omega}_{1}$.  By introducing an additional term 
proportional to the first class constraints 
$\tilde{\Omega}_{2}$ into $\tilde{H}$, we then obtain an equivalent first class 
Hamiltonian 
\beq
\tilde{H}^{\prime}=\tilde{H}+\int{\rm d}^{3}x~\pi_{\theta}\tilde{\Omega}_{2}
\label{htildeprime}
\eeq
which naturally generates the Gauss law constraint 
\beq
\{\tilde{\Omega}_{1},\tilde{H}^{\prime}\}=2\tilde{\Omega}_{2},~~~
\{\tilde{\Omega}_{2},\tilde{H}^{\prime}\}=0.
\eeq
One notes here that $\tilde{H}$ and $\tilde{H}^{\prime}$ act in the same way 
on physical states, which are annihilated by the first-class constraints. 
Similarly, the equations of motion for observables remain unaffected by the 
additional term in $\tilde{H}^{\prime}$.  Furthermore, on the zero locus of the 
constraints $(\theta,\pi_{\theta})$, our first class system 
is exactly reduced to the original second class one.

Next, we consider the Poisson brackets of the fields in the extended phase
space $\tilde{{\cal F}}$ and identify the Dirac brackets by taking the
vanishing limit of auxiliary fields. After some algebraic manipulation 
starting from Eq. (\ref{pitilde}), one can obtain the commutators 
\begin{eqnarray}
\{\tilde{\sigma}(x),\tilde{\sigma}(y)\}&=&\{\tilde{\pi}_{\sigma}(x),
\tilde{\pi}_{\sigma}(y)\}=0,
\nonumber\\
\{\tilde{\sigma}(x),\tilde{\pi}_{\sigma}(y)\}&=&\left
(1-\frac{\tilde{\sigma}^{2}}{\tilde{\sigma}^{2}
+\tilde{\pi}_{k}\tilde{\pi}_{k}}\right)\delta(x-y),  
\nonumber\\
\{\tilde{\pi}_{a}(x),\tilde{\pi}_{b}(y)\}&=&0,  \nonumber\\
\{\tilde{\pi}_{a}(x),\tilde{\pi}_{\pi}^{b}(y)\}&=&
\left(\delta_{ab}-\frac{\tilde{\pi}_{i}\tilde{\pi}_{j}}{\tilde{\sigma}^{2}
+\tilde{\pi}_{k}\tilde{\pi}_{k}}\delta_{ai}\delta_{bj}\right)\delta(x-y),  
\nonumber\\
\{\tilde{\pi}_{\pi}^{a}(x),\tilde{\pi}_{\pi}^{b}(y)\}&=&
\frac{1}{\tilde{\sigma}^{2}+\tilde{\pi}_{k}\tilde{\pi}_{k}}\left(\tilde{\pi}_{j}
\tilde{\pi}_{\pi}^{i}-\tilde{\pi}_{i}\tilde{\pi}_{\pi}^{j}\right)
\delta_{ai}\delta_{bj}\delta(x-y),  
\nonumber\\
\{\tilde{\psi}(x),\tilde{\psi}(y)\}&=&
\{\tilde{\pi}_{\psi}^{\dagger}(x),\tilde{\pi}_{\psi}^{\dagger}(y)\}=0,
\nonumber\\
\{\tilde{\psi}(x),\tilde{\pi}_{\psi}^{\dagger}(y)\}&=&\delta(x-y).
\label{commst}
\end{eqnarray}
One notes here that on the zero locus of the constraints $(\theta,\pi_{\theta})$, 
the above Poisson brackets in the extended phase space exactly reproduce the corresponding Dirac brackets (\ref{commdirac}).  It is also noteworthy that the Poisson brackets of the fields $\tilde{{\cal F}}$ in Eq. (\ref{commst}) have exactly the same form as those of the Dirac brackets of the field ${\cal F}$ to yield 
$\{\tilde{A},\tilde{B}\}=\{A,B\}_{D}|_{A\rightarrow \tilde{A},B\rightarrow 
\tilde{B}}$.  On the other hand, this kind of situation happens again 
when one considers the first-class
constraints (\ref{1stconst}). More precisely, these first-class constraints
in the extended phase space can be rewritten as 
\beq
\tilde{\Omega}_{1}=\tilde{\sigma}^{2}+\tilde{\pi}_{i}\tilde{\pi}_{i}-f_{\pi}^{2},~~~
\tilde{\Omega}_{2}=\tilde{\sigma}\tilde{\pi}_{\sigma}
+\tilde{\pi}_{i}\tilde{\pi}_{\pi}^{i},  
\label{oott}
\eeq
which are form-invariant with respect to the second-class constraints (\ref
{c1}) and (\ref{const22}).

%%%%%%%%%%%%%%%%%%%%%%%%%%%%%%%%%%%%%%%%%%%%%%%%%%%%%%%%%%%%%%%%%%%%%%%%%%%

\section{BRST symmetries}
\setcounter{equation}{0}
\renewcommand{\theequation}{\arabic{section}.\arabic{equation}}

%%%%%%%%%%%%%%%%%%%%%%%%%%%%%%%%%%%%%%%%%%%%%%%%%%%%%%%%%%%%%%%%%%%%%%%%%%%

In this section, we will obtain the BRST invariant Lagrangian in the framework 
of the BFV formalism \cite{fradkin75} which is 
applicable to theories with the first class constraints by introducing two 
canonical sets of ghosts and anti-ghosts together with auxiliary fields 
$({\cal C}^{i},\bar{{\cal P}}_{i})$, $({\cal P}^{i},\bar{{\cal C}}_{i})$, 
$(N^{i},B_{i})$, $(i=1,2)$ which satisfy the super-Poisson algebra
\beq
\{{\cal C}^{i}(x),\bar{{\cal P}}_{j}(y)\}
=\{{\cal P}^{i}(x),\bar{{\cal C}}_{j}(y)\}
=\{N^{i}(x),B_{j}(y)\}=\delta_{j}^{i}\delta(x-y).
\eeq
Here the super-Poisson bracket is defined as
\beq
\{A,B\}=\frac{\delta A}{\delta q}|_{r}\frac{\delta B}{\delta p}|_{l}
-(-1)^{\eta_{A}\eta_{B}}\frac{\delta B}{\delta q}|_{r}\frac{\delta A} {%
\delta p}|_{l}
\eeq
where $\eta_{A}$ denotes the number of fermions called ghost number in $A$
and the subscript $r$ and $l$ right and left derivatives.

In this phenomenological SU(3) linear sigma model, the nilpotent BRST charge $Q$, the fermionic
gauge fixing function $\Psi$ and the BRST invariant minimal Hamiltonian $%
H_{m}$ are given by
\begin{eqnarray}
Q&=&\int{\rm d}^{3}x~({\cal C}^{i}\tilde{\Omega}_{i}+{\cal P}^{i}B_{i}),\nonumber\\
\Psi&=&\int{\rm d}^{3}x~(\bar{{\cal C}}_{i}\chi^{i}+\bar{{\cal P}}_{i}N^{i}),  \nonumber \\
H_{m}&=&\tilde{H}^{\prime}-\int{\rm d}^{3}x~2{\cal C}^{1}
\bar{{\cal P}}_{2}
\end{eqnarray}
which satisfy the relations $\{Q,H_{m}\}=0$, $Q^{2}=\{Q,Q\}=0$, $\{\{\Psi,Q\},Q\}=0$.
The effective quantum Lagrangian is then described as
\begin{equation}
L_{eff}=\int{\rm d}^{3}x~(\pi_{\sigma}\dot{\sigma}+\pi_{\pi}^{a}\dot{\pi}_{a}
+\pi_{\psi}^{\dagger}\dot{\psi}+\pi_{\theta}\dot{\theta}+B_{2}\dot{N}^{2}+
\bar{{\cal P}}_{i}\dot{{\cal C}}^{i}+\bar{{\cal C}}_{2} \dot{{\cal P}}^{2})-H_{tot}
\end{equation}
with $H_{tot}=H_{m}-\{Q,\Psi\}$. Here $B_{1}\dot{N}^{1} +\bar{{\cal C}}_{1}%
\dot{{\cal P}}^{1}=\{Q,\bar{{\cal C}}_{1} \dot{N}^{1}\}$ terms are
suppressed by replacing $\chi^{1}$ with $\chi^{1} +\dot{N}^{1}$.

Now we choose the unitary gauge
\begin{equation}
\chi^{1}=\Omega_{1},~~~\chi^{2}=\Omega_{2}
\end{equation}
and perform the path integration over the fields $B_{1}$, $N^{1}$, $\bar{%
{\cal C}}_{1}$, ${\cal P}^{1}$, $\bar{{\cal P}}_{1}$ and ${\cal C}^{1}$, by
using the equations of motion, to yield the effective Lagrangian of the form
\begin{eqnarray}
L_{eff}&=&\int{\rm d}^{3}x~\left[\pi_{\sigma}\dot{\sigma}+\pi_{\pi}^{a}\dot{\pi}_{a}
+\pi_{\psi}^{\dagger}\dot{\psi}+\pi_{\theta}\dot{\theta}+B_{2}\dot{N}^{2}+
\bar{{\cal P}}_{2}\dot{{\cal C}}^{2}+\bar{{\cal C}}_{2} \dot{{\cal P}}^{2}
\right.
\nonumber\\
& &\left.-\frac{R}{2}\left(\left(\pi_{\sigma}-\sigma\pi_{\theta}\right)^{2}
+\left(\pi_{\pi}^{i}-\pi_{i}\pi_{\theta}\right)^{2}\right)
-\frac{1}{2}\pi_{\pi}^{\bar{a}}\pi_{\pi}^{\bar{a}}
\right.\nonumber\\
& &\left.
-\frac{1}{2R}\left((\partial_{i}\sigma)^{2}+(\partial_{i}\pi_{k})^{2}\right)
-\frac{1}{2}(\partial_{i}\pi_{\bar{a}})^{2}\right.\nonumber\\
& &\left.
-\frac{1}{2}\mu_{0}^{2}\left(\frac{1}{R}(\sigma^{2}+\pi_{i}\pi_{i})
+\pi_{\bar{a}}\pi_{\bar{a}}\right)
-\frac{1}{4}\lambda_{0}\left(\frac{1}{R}(\sigma^{2}+\pi_{i}\pi_{i})
+\pi_{\bar{a}}\pi_{\bar{a}}\right)^{2}
\right.\nonumber\\
& &\left.
-\bar{\psi}i\gamma^{i}\partial_{i}\psi
-g_{0}\bar{\psi}\frac{1}{\sqrt{2R}}(\sigma+i\gamma_{5}\pi_{i}\tau_{i})\psi
-g_{0}\bar{\psi}\frac{1}{\sqrt{2}}i\gamma_{5}\pi_{\bar{a}}\lambda_{\bar{a}}\psi
\right.\nonumber\\
& &\left.
+\left(\sigma\pi_{\sigma}+\pi_{i}\pi_{\pi}^{i}-(\sigma^{2}+\pi_{k}\pi_{k})\pi_{\theta}\right)
(-\pi_{\theta}+N)
-2(\sigma^{2}+\pi_{k}\pi_{k})\pi_{\theta}{\cal C}\bar{\cal C}
\right.\nonumber\\
& &\left.
+(\sigma\pi_{\sigma}+\pi_{i}\pi_{\pi}^{i})B+\bar{\cal P}{\cal P}-\mu_{\pi}\rho_{\pi}-\mu_{K}\rho_{K}
\right]
\end{eqnarray}
with redefinitions: $N\equiv N^{2}$, $B\equiv B_{2}$, $\bar{{\cal C}}\equiv
\bar{{\cal C}}_{2}$, ${\cal C}\equiv {\cal C}^{2}$, $\bar{{\cal P}}\equiv
\bar{{\cal P}}_{2}$, ${\cal P}\equiv {\cal P}_{2}$ and 
$R=(\sigma^{2}+\pi_{i}\pi_{i})/(\sigma^{2}+\pi_{i}\pi_{i}+2\theta)$.

Next, using the variations with respect to $\pi_{\sigma}$, $\pi_{\pi}^{a}$, $\pi_{\theta}$, 
${\cal P}$ and $\bar{{\cal P}}$, one obtain the relations
\begin{eqnarray}
\dot{\sigma}&=&(\pi_{\sigma}-\sigma\pi_{\theta})R+\sigma(\pi_{\theta}-N-B)
\nonumber\\
\dot{\pi}_{i}&=&(\pi_{\pi}^{i}-\pi_{i}\pi_{\theta})R+\pi_{i}(\pi_{\theta}-N-B)
+\mu_{\pi}(\pi_{1}\delta_{i}^{2}-\pi_{2}\delta_{i}^{1})
\nonumber\\
\dot{\pi}_{\bar{a}}&=&\pi_{\pi}^{\bar{a}}
+\mu_{K}(\pi_{4}\delta_{\bar{a}}^{5}-\pi_{5}\delta_{\bar{a}}^{4})
\nonumber\\
\dot{\theta}&=&-\sigma(\pi_{\sigma}-\sigma\pi_{\theta})R
-\pi_{i}(\pi_{\pi}^{i}-\pi_{i}\pi_{\theta})R
+(\sigma^{2}+\pi_{i}\pi_{i})(-2\pi_{\theta}+N+2{\cal C}\bar{\cal C})
\nonumber\\
& &+\sigma\pi_{\sigma}+\pi_{i}\pi_{\pi}^{i}
\nonumber \\
{\cal P}&=&-\dot{{\cal C}},~~~~~\bar{{\cal P}}=\dot{\bar{{\cal C}}}
\end{eqnarray}
to yield the effective Lagrangian
\begin{eqnarray}
L_{eff}&=&\int{\rm d}^{3}x~\left[
\frac{1}{2R}\left((\partial_{\mu}\sigma)^{2}+(\partial_{\mu}\pi_{k})^{2}\right)
+\frac{1}{2}(\partial_{\mu}\pi_{\bar{a}})^{2}
\right.\nonumber\\
& &\left.
-\frac{1}{2}\mu_{0}^{2}\left(\frac{1}{R}(\sigma^{2}+\pi_{i}\pi_{i})
+\pi_{\bar{a}}\pi_{\bar{a}}\right)
-\frac{1}{4}\lambda_{0}\left(\frac{1}{R}(\sigma^{2}+\pi_{i}\pi_{i})
+\pi_{\bar{a}}\pi_{\bar{a}}\right)^{2}
\right.\nonumber\\
& &\left.
+\bar{\psi}i\gamma^{\mu}\partial_{\mu}\psi
-g_{0}\bar{\psi}\frac{1}{\sqrt{2R}}(\sigma+i\gamma_{5}\pi_{i}\tau_{i})\psi
-g_{0}\bar{\psi}\frac{1}{\sqrt{2}}i\gamma_{5}\pi_{\bar{a}}\lambda_{\bar{a}}\psi
\right.\nonumber\\
& &\left.
-\frac{1}{2R}(\sigma^{2}+\pi_{i}\pi_{i})\left(\frac{\dot{\theta}}
{\sigma^{2}+\pi_{i}\pi_{i}}+(B+2\bar{\cal C}{\cal C})R\right)^{2}
\right.\nonumber\\
& &\left.
+\frac{1}{R}(B+N)\left(-\dot{\theta}+(\sigma^{2}+\pi_{i}\pi_{i})
\left(\frac{\dot{\theta}}{\sigma^{2}+\pi_{i}\pi_{i}}+(B+2\bar{\cal C}{\cal C})R
\right)\right)
\right.\nonumber\\
& &\left.
+B\dot{N}-\bar{\cal C}\partial_{0}^{2}{\cal C}
-\mu_{\pi}\rho_{\pi}-\mu_{K}\rho_{K}
\right].
\end{eqnarray}

Finally, with the identification 
$N=-B+\frac{\dot{\theta}}{\sigma^{2}+\pi_{i}\pi_{i}}$, one can arrive at the 
BRST invariant Lagrangian 
\begin{eqnarray}
L_{eff}&=&\int{\rm d}^{3}x~\left[
\frac{1}{2}((\partial_{\mu}\sigma)^{2}+(\partial_{\mu}\pi_{k})^{2})\frac{\sigma^{2}+\pi_{i}\pi_{i}+2\theta}
{\sigma^{2}+\pi_{i}\pi_{i}}
+\frac{1}{2}(\partial_{\mu}\pi_{\bar{a}})^{2}
\right.\nonumber\\
& &\left.
-\frac{1}{2}\mu_{0}^{2}(\sigma^{2}+\pi_{i}\pi_{i}+2\theta+\pi_{\bar{a}}\pi_{\bar{a}})
-\frac{1}{4}\lambda_{0}(\sigma^{2}+\pi_{i}\pi_{i}+2\theta+\pi_{\bar{a}}\pi_{\bar{a}})^{2}
\right.\nonumber\\
& &\left.
+\bar{\psi}i\gamma^{i}\partial_{i}\psi
-g_{0}\bar{\psi}\frac{1}{\sqrt{2}}(\sigma+i\gamma_{5}\pi_{i}\tau_{i})\psi
\left(\frac{\sigma^{2}+\pi_{i}\pi_{i}+2\theta}
{\sigma^{2}+\pi_{i}\pi_{i}}\right)^{1/2}
\right.\nonumber\\
& &\left.
-g_{0}\bar{\psi}\frac{1}{\sqrt{2}}i\gamma_{5}\pi_{\bar{a}}\lambda_{\bar{a}}\psi
-\frac{1}{2}\frac{\sigma^{2}+\pi_{i}\pi_{i}+2\theta}{(\sigma^{2}+\pi_{i}\pi_{i})^{2}}
\dot{\theta}^{2}-\frac{\dot{B}\dot{\theta}}{\sigma^{2}+\pi_{i}\pi_{i}}
\right.\nonumber\\
& &\left.
-\frac{1}{2}\frac{(\sigma^{2}+\pi_{i}\pi_{i})^{2}}{\sigma^{2}+\pi_{i}\pi_{i}+2\theta}
(B+2\bar{\cal C}{\cal C})^{2}
-\bar{\cal C}\partial_{0}^{2}{\cal C}
-\mu_{\pi}\rho_{\pi}-\mu_{K}\rho_{K}
\right].
\end{eqnarray}
which is invariant under the BRST transformation
\begin{eqnarray}
\delta_{B}\sigma&=&\lambda\sigma{\cal C},~~~ 
\delta_{B}\pi_{i}=\lambda\pi_{i}{\cal C},~~~ 
\delta_{B}\pi_{\bar{a}}=0,~~~
\delta_{B}\psi=\lambda\psi{\cal C},
\nonumber\\ 
\delta_{B}\theta&=&-\lambda(\sigma^{2}+\pi_{i}\pi_{i}){\cal C},~~~ 
\delta_{B}\bar{{\cal C}}=-\lambda B,~~~ \delta_{B}{\cal C}=\delta_{B}B=0.
\end{eqnarray}

%%%%%%%%%%%%%%%%%%%%%%%%%%%%%%%%%%%%%%%%%%%%%%%%%%%%%%%%%%%%%%%%%%%%%%%%%%%

\section{Phenomenology and discussions}
\setcounter{equation}{0}
\renewcommand{\theequation}{\arabic{section}.\arabic{equation}}

%%%%%%%%%%%%%%%%%%%%%%%%%%%%%%%%%%%%%%%%%%%%%%%%%%%%%%%%%%%%%%%%%%%%%%%%%%%

In this section, to discuss phenomenological aspects, we exploit the 
first-class constraints $\tilde{\Omega}_{i}=0$ in Eq. (\ref{1stconst}) to 
the Hamiltonian (\ref{htildeprime}) to obtain the relation
\bea
& &\frac12\left(\left(\pi_{\sigma}-\sigma\pi_{\theta}\right)^{2}
+\left(\pi_{\pi}^{i}-\pi_{i}\pi_{\theta}\right)^{2}\right)
\frac{\sigma^{2}+\pi_{k}\pi_{k}}
{\sigma^{2}+\pi_{k}\pi_{k}+2\theta}
\nonumber\\
& &=\frac{1}{2f_{\pi}^{2}}\left((\sigma^{2}+\pi_{i}\pi_{i})\left(\pi_{\sigma}^{2}
+\pi_{\pi}^{i}\pi_{\pi}^{i}\right)-\left(\sigma\pi_{\sigma}
+\pi_{i}\pi_{\pi}^{i}\right)^{2}\right).
\label{convertion}
\eea
Following the symmetrization procedure, we then obtain a Hamiltonian of the 
form\begin{eqnarray}
\tilde{H}&=&\int {\rm d}^{3}x~\left[
\frac12\left(\pi_{\sigma}^{2}
+\pi_{\pi}^{i}\pi_{\pi}^{i}+1\right)
+\frac{1}{2}\pi_{\pi}^{\bar{a}}\pi_{\pi}^{\bar{a}}
+\frac12((\partial_{i}\sigma)^{2}+(\partial_{i}\pi_{k})^{2})
\right.\nonumber\\
& &\left.
+\frac{1}{2}(\partial_{i}\pi_{\bar{a}})^{2}
+\frac12\mu_{0}^{2}(\sigma^{2}+\pi_{i}\pi_{i})
+\frac{1}{2}\mu_{0}^{2}\pi_{\bar{a}}\pi_{\bar{a}}
+\frac{1}{4}\lambda_{0}(\sigma^{2}+\pi_{i}\pi_{i})^{2}
\right.\nonumber\\
& &\left.
+\frac{1}{4}\lambda_{0}(\pi_{\bar{a}}\pi_{\bar{a}})^{2}
+\frac{1}{2}\lambda_{0}\pi_{\bar{a}}\pi_{\bar{a}}(\sigma^{2}+\pi_{i}\pi_{i})
+\mu_{\pi}\rho_{\pi}+\mu_{K}\rho_{K}
+\bar{\psi}i\gamma^{i}\partial_{i}\psi
\right.\nonumber\\
& &\left.
+g_{0}\bar{\psi}\frac{1}{\sqrt{2}}
(\sigma+i\gamma_{5}\pi_{i}\tau_{i})\psi
+g_{0}\bar{\psi}\frac{1}{\sqrt{2}}i\gamma_{5}\pi_{\bar{a}}\lambda_{\bar{a}}\psi
\right].
\label{hctwyle}
\end{eqnarray}
Here one notes that a Weyl ordering correction $\frac{1}{2}$ in the first line of 
Eq. (\ref{hctwyle}) originates from the improved Dirac scheme associated with the 
geometric constraint (\ref{c1}).  Moreover, this correction comes only with the 
kinetic terms, without any dependence on the potential terms.  

Now we define mean fields for the Goldstone boson fields as\footnote{
Here we ignore the eta fields for simplicity.}
\beq
\begin{array}{lll}
\langle\sigma\rangle=\sigma, 
&\langle\pi^{\pm}\rangle=\pi^{\pm}, 
&\langle K^{\pm}\rangle=K^{\pm},\\ 
\langle\pi_{\sigma}\rangle=p_{\sigma}, 
&\langle\pi_{\pi^{\pm}}\rangle=p_{\pi^{\pm}}, 
&\langle\pi_{K^{\pm}}\rangle=p_{K^{\pm}},\\
{\rm others}=0, 
& 
&
\end{array}
\eeq
where 
\bea
\pi^{\pm}&=&\frac{1}{\sqrt{2}}(\pi_{1}\mp i\pi_{2}),
~~\pi^{0}=\pi_{3},~~
K^{\pm}=\frac{1}{\sqrt{2}}(\pi_{4}\mp i\pi_{5}),\nonumber\\
K^{0}&=&\frac{1}{\sqrt{2}}(\pi_{6}-i\pi_{7}),~~
\bar{K}^{0}=\frac{1}{\sqrt{2}}(\pi_{6}+i\pi_{7}),
\eea
and we have similar relations for the momenta fields.  
We then finally arrive at the energy spectrum of the form 
$\langle\tilde{H}\rangle=\int {\rm d}^{3}x~\varepsilon$ with
\bea
\varepsilon&=&
\frac12 p_{\sigma}^{2}+p_{\pi^{+}}p_{\pi^{-}}+p_{K^{+}}p_{K^{-}}
+\frac12(\partial_{i}\sigma)^{2}+\partial_{i}\pi^{+}\partial_{i}\pi^{-}
+\partial_{i}K^{+}\partial_{i}K^{-}\nonumber\\
& &+\frac12\mu_{0}^{2}(\sigma^{2}+2\pi^{+}\pi^{-}+2K^{+}K^{-})
+\frac{1}{4}\lambda_{0}(\sigma^{2}+2\pi^{+}\pi^{-}+2K^{+}K^{-})^{2}
\nonumber\\
& &+\bar{\psi}i\gamma^{i}\partial_{i}\psi
+g_{0}\bar{\psi}\left(\frac{1}{\sqrt{2}}\sigma+i\gamma_{5}
(\tau^{+}\pi^{-}+\tau^{-}\pi^{+}+\lambda^{+}K^{-}+\lambda^{-}K^{+})\right)\psi
\nonumber\\
& &+i\mu_{\pi}(\pi^{-}p_{\pi^{+}}-\pi^{+} p_{\pi^{-}})
+i\mu_{K}(K^{-}p_{K^{+}}-K^{+}p_{K^{-}})
\nonumber\\
& &+\psi^{\dagger}\left(\mu_{\pi}(\hat{Q}_{u}+\hat{Q}_{d})+\mu_{K}\hat{Q}_{s}\right)\psi
+\frac12 
\label{hctwyle2}
\eea
where
\beq
\tau^{\pm}=\frac12(\tau_{1}\mp i\tau_{2}),~~~\tau^{0}=\tau_{3},~~~
\lambda^{\pm}=\frac12(\lambda_{4}\mp i\lambda_{5}).
\eeq
Using the variations with respect to $p_{\sigma}$, $p_{\pi^{\pm}}$ and 
$p_{K^{\pm}}$, we obtain the relations
\beq
p_{\sigma}=0,~~~~
p_{\pi^{\pm}}=\pm i\mu_{\pi}\pi^{\pm},~~~~
p_{K^{\pm}}=\pm i\mu_{K}K^{\pm}
\label{eqnofmot2}
\eeq
to yield
\beq
\dot{\sigma}=0,~~~\dot{\pi}^{\pm}=\pm i\mu_{\pi}\pi^{\pm},~~~~\dot{K^{\pm}}=\pm i\mu_{K}K^{\pm}.
\label{dotdot}
\eeq

Substituting the Eq. (\ref{eqnofmot2}) into the energy spectrum (\ref{hctwyle2}) 
and ignoring the irrelevant term $(\partial_{i}\sigma)^{2}$, we finally arrive at 
\bea
\varepsilon&=&\partial_{i}\pi^{+}\partial_{i}\pi^{-}
+\partial_{i}K^{+}\partial_{i}K^{-}
+\frac12\mu_{0}^{2}\sigma^{2}
-(\mu_{\pi}^{2}-\mu_{0}^{2})\pi^{+}\pi^{-}
-(\mu_{K}^{2}-\mu_{0}^{2})K^{+}K^{-}
\nonumber\\
& &+\frac{1}{4}\lambda_{0}(\sigma^{2}+2\pi^{+}\pi^{-} 
+2\pi^{+}\pi^{-}+2K^{+}K^{-})^{2}+\bar{\psi}i\gamma^{i}\partial_{i}\psi
\nonumber\\
& &+g_{0}\bar{\psi}\left(\frac{1}{\sqrt{2}}\sigma+i\gamma_{5}
(\tau^{+}\pi^{-}+\tau^{-}\pi^{+}+\lambda^{+}K^{-}+\lambda^{-}K^{+})\right)\psi
\nonumber\\
& &+\psi^{\dagger}\left(\mu_{\pi}(\hat{Q}_{u}+\hat{Q}_{d})+\mu_{K}\hat{Q}_{s}
\right)\psi+\frac12,
\label{hctwyle3}
\eea
which still respects the SU(3) flavor symmetry.

%%%%%%%%%%%%%%%%%%%%%%%%%%%%%%%%%%%%%%%%%%%%%%%%%%%%%%%%%%%%%%%%%%%%%%%%%%

%\section{Conclusion}

%%%%%%%%%%%%%%%%%%%%%%%%%%%%%%%%%%%%%%%%%%%%%%%%%%%%%%%%%%%%%%%%%%%%%%%%%

In summary, we constructed the SU(3) linear 
sigma model by introducing a novel matrix for the Goldstone bosons 
which satisfy geometrical second-class constraints.  Following the 
improved Dirac method, we also constructed first-class physical fields and, 
in terms of them, we directly obtained a first-class Hamiltonian which 
is consistent with the Hamiltonian with the original fields and auxiliary fields. 
The Poisson brackets of the first-class physical fields are also constructed and 
these Poisson brackets are shown to reproduce the corresponding Dirac brackets 
in the limit of vanishing auxiliary fields.  Exploiting the first-class 
Hamiltonian, we constructed the BRST invariant effective Lagrangian and its 
corresponding BRST transformation rules in this phenomenological SU(3) linear sigma model.  
Finally, defining the mean fields for the Goldstone bosons fields, we obtained 
the energy spectrum of the corresponding pion and kaon condensates.  However this energy 
spectrum still possesses the SU(3) flavor symmetry.  Through further investigation, it 
will be interesting to study the flavor symmetry breaking effects in the framework of this 
SU(3) linear sigma model to predict the more realistic kaon condensation.    

\vskip 1.0cm 
STH would like to thank the hospitality of the Institute of Physics and Applied 
Physics, Yonsei University where a part of this work has been carried out.  
We would like to thank S. Ando, K. Kubodera, F. Myhrer, Y. Oh, Y.J. Park and M. Rho 
for helpful discussions.  STH acknowledges financial support in part from 
the Korea Research Foundation, Grant No. KRF-2001-DP0083.  SHL acknowledges 
financial support from KOSEF under grant No. 1999-2-111-005-5, and 
the Korean Ministry of Education under Grant No. 2000-2-0689.

\end{document}